\documentstyle[12pt,epsf,colordvi,aaspp4]{article}

\begin{document}

\title{Good views of the Galaxy}

\author{Hiranya V. Peiris}
\affil{Princeton University Observatory, Peyton Hall, Princeton, NJ 08544;   
	hiranya@astro.princeton.edu}

\begin{abstract} 

Fitting Galactic structure models to star counts only provides useful
information about the Galaxy in some directions. In this paper, we
investigate the use of $\chi^2$ goodness-of-fit tests to discriminate
between degenerate Galactic structure models, and the implications of this
technique for the Galactic spheroid and thick disk components. The axis
ratio of the Galactic spheroid and the normalization of spheroid stars
with respect to disk stars introduce a degenerate effect which means that
Galactic structure models with certain combinations of these parameters
are indistinguishable from each other in most directions. We present an
analysis of the optimal directions in which these degeneracies can be
lifted. Poisson and magnitude errors are taken into account, and an
attempt is made to place an upper limit on the systematic error due to
separation of spheroid stars from thick/old disk stars. We find that the
magnitude range $20 < V < 21$ is the best for lifting most degeneracies,
and present the optimal combinations of directions using which this can be
achieved.  We also give directions in which the signature of the presence
of a Galactic thick disk can be most readily identified, and the
directions in which contamination from a thick disk can be minimized. It
is hoped that forthcoming data from large-scale sky surveys would reveal
much about the structure of our Galaxy using star count techniques.

\end{abstract}



\section{Introduction} \label{intro}

Since we observe our Galaxy from within it, we must use indirect tools
such as star counts to probe its structure. The success of this method
relies on assuming that the components of the Milky Way Galaxy have stars
distributed similarly to those of galaxies of the same Hubble type, and
comparing models with observations only in regions where little
obscuration is known to be present. Observed ``star counts'' (i.e. the
distribution of number counts of stars in apparent magnitude and color
within a given area/direction in the sky) are then used to determine
parameters in distribution functions whose overall shapes are assumed
known.

Within the past two decades, Galactic structure models of varying degrees
of complexity have been developed (Bahcall \& Soneira, 1980, 1981, 1984,
\cite{gil84}, Robin and Cr\'ez\'e 1986a, 1986b, \cite{rei93}).  However,
during the same period, relatively few sets of observational star count
data have been published.

Galactic structure models, for a given set of structure parameters,
typically predict differential number counts (stars / apparent magnitude
bin / area) and frequency-color distributions (stars / color bin / area)
for a specified direction. Because the Galaxy consists of at least two
components, obeying different density laws and being characterized by
different stellar populations, these distributions change with the color
range, magnitude range and direction under consideration. It is important
to match both the shapes of these distributions and the absolute numbers
of stars with observations.

Existing models predict differential number counts reasonably well, to
faint magnitudes of $V \sim 21$, but recent studies suggest that agreement
between observed and predicted frequency-color distributions breaks down
at faint magnitudes below $V \sim 19$ (\cite{rei93}). ``Standard''
Galactic structure models appear to overestimate the contribution of
(blue) spheroid stars substantially, and the contribution of (red) disk
stars is correspondingly underestimated in order to match the total number
counts. Contrary to prediction, observations show that {\it disk} stars
are the majority population to at least $V \sim 21$. However, the small
sizes of fields in which data are available mean that comparisons in
frequency-color space are complicated by uncertainties due to small-number
statistics.

A further hindrance to the determination of distribution function
parameters arises due to the fact that data are only available in a small
number of directions. This presents a problem in trying to fit model
parameters which cause degenerate effects in the number counts. For
instance, increasing the normalization of the Galactic spheroid and
decreasing its axis ratio \slantfrac{b}{a} (or vice versa) represents a
degeneracy. This means that in most directions, one cannot distinguish
between models with a flattened spheroid plus a low normalization ratio of
spheroid stars to disk stars, and those with a high axis ratio plus a high
normalization.

We anticipate the availability of star count data from large-scale, deep
sky surveys in the near future (see \S~\ref{dirn} and \S~\ref{finish})
which would solve the problem of excessive statistical noise, and provide
data in almost any direction we desire. In this paper, we investigate the
possibility of tackling the degeneracy problem by attempting to find two
or more directions in which $\chi^2$ goodness-of-fit tests can distinguish
between spheroid parameter combinations that yield degenerate results in
general.

A degenerate situation is defined as one in which, in a general direction,
there is no significant difference in results given by a model where a
pair of spheroid parameters are characterized by values $(p, q)$ and a
model characterized by values $(r, s)$. This means that one cannot learn
very much about these spheroid parameters even if these models fit the
spheroidal star counts in that direction with reasonable accuracy. The
degenerate parameters we examine here are the spheroid axis ratio and the
normalization of spheroid stars to disk stars in the solar neighborhood.

By using $\chi^2$ tests with a specified set of directions as ``bins"  
(see e.g. \cite{pre92}), we have investigated the optimal directions for
breaking this degeneracy. These tests are based on fitting the total
number of spheroid stars in each direction ``bin'', for a given set of
directions and a particular magnitude range. Therefore, both the observed
number count and color information is used, since the spheroid counts have
to be separated from the total counts using the frequency-color
distribution, as explained in \S~\ref{seperr}.

\S~\ref{model} describes the Galactic structure models we have used in
these tests. Details of the directions used in the analysis are found in
\S~\ref{dirn}. A description of the $\chi^2$ tests, along with our
results, are presented in \S~\ref{chisq}. We conclude, in \S~\ref{finish},
that there are two or more directions in which most ``degenerate" model
pairs can be distinguished. We find the magnitude range $20 < V < 21$ to
be the most useful for this purpose. We also make suggestions for the best
directions in which to separate spheroidal stars from old- and thick-disk
stars using a frequency-color distribution diagram, and conversely, the
directions in which the presence of a thick disk would be most clearly
observed.


\section{The Models} \label{model}

While it is our intention to illustrate the use of this technique in
general, for specificity we will use the model that is usually referred to
in the literature as the Bahcall-Soneira (B\&S) model. We have taken the
realization of the B\&S model in the form of the Export Code that is
available on the Internet
(http://www.sns.ias.edu/\verb+~+jnb/Html/galaxy.html). This model is
described in Bahcall (1986). We have added a few refinements as described
below.

The disk luminosity function (LF) has been modified between $9.5 < M_V
\leq 18.0$ to match the LF recently derived from {\it Hubble\ Space\
Telescope} ({\it HST}) M dwarfs (Figure 2, \cite{gou97}). A ``sech$^2$"  
model has been used for the vertical density distribution function of main
sequence disk stars in the same absolute magnitude range, as given in Eq.
3.2 of the same work. The HST LF was derived using two different methods:
a maximum likelihood (ML) fit that takes into account both measurement
errors and Malmquist bias, and a naive binning method where the total
number of stars in a given magnitude bin is divided by the effective
volume integrated over that bin. This binning method takes into account
neither Malmquist bias nor observational error. The two alternative
derivations agree quite well over most of the LF, but at the faint end
($13.5 \leq M_V \leq 18.5$) the ML procedure could by affected by a
statistical fluctuation. Because Poisson errors are potentially a much
more serious problem at the faint end than Malmquist bias (since the HST
survey extends to the ``top'' of the disk), we choose to use the form of
the LF derived using the binning method.

The spheroid LF has been modified between $7.5 \leq M_V \leq 13.5$ to
match another LF derived from {\it HST} star counts (Figure 3,
\cite{gou98}). Note that this LF differs in its zero point by a factor of
$\sim 2$ with studies of kinematically-selected local spheroid subdwarfs
(\cite{dah95}).

\begin{figure} \plotone{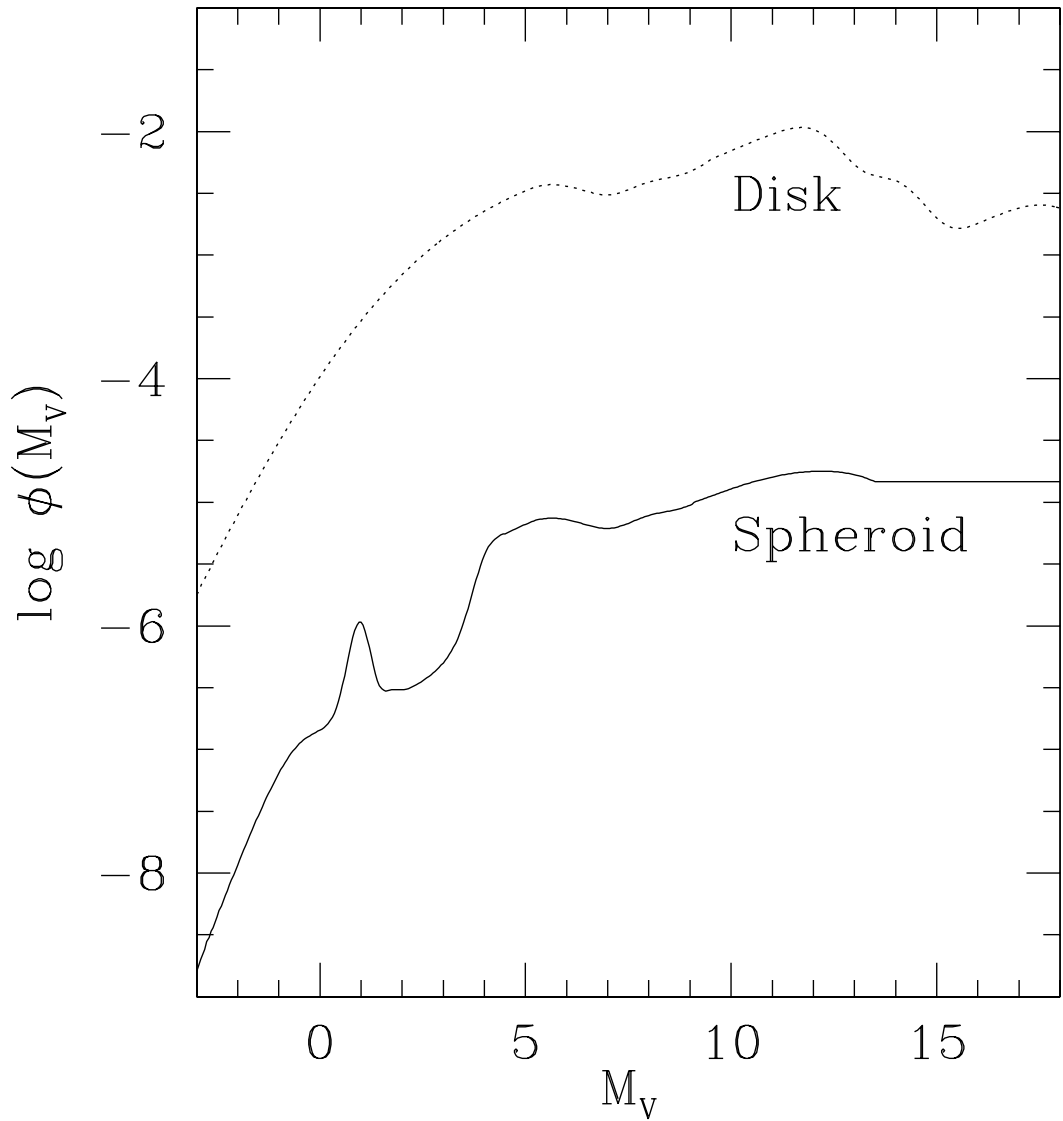} \caption{The adopted disk and spheroid
luminosity functions (LFs). The LFs adopted are the same form those
described in Bahcall (1986) except for the following refinements: the disk
LF for $M_V > 9.5$ has been modified to match smoothly the LF derived from
$HST$ star counts given in Figure 2 of Gould et al.\ (1997) and the
spheroid LF for $M_V > 7.5$ has been modified to match the LF given in
Figure 3 of Gould et al.\ (1998), also derived from $HST$ data.
\label{fig-lf}} \end{figure}

The disk and spheroid LFs used are shown in Figure~\ref{fig-lf}. In
addition to the spheroidal parameters whose values are described in
\S~\ref{chisq}, the color-magnitude diagram (CMD) of M13 was adopted for
the spheroid CMD.

\placefigure{fig-lf}

In addition to this two-component model, a three-component model was
created with the addition of a thick disk with a scale height of 1200 pc,
a 47 Tuc-like CMD, a disk-like LF, and a normalization to disk counts at
the solar neighborhood of 2\%. This was used in analyzing the separation
error in identifying spheroid counts from disk/thick-disk counts using
frequency-color diagrams, as described in \S~\ref{seperr}.

The density laws used for the old disk, the spheroid and the thick disk
are summarized in Table~\ref{tbl-dist}, which also gives the
normalizations used.

\placetable{tbl-dist}

\section{The Directions} \label{dirn}

We selected a representative set of directions in which to carry out the
analysis, which are shown in Table~\ref{tbl-dir}.

\placetable{tbl-dir}

Directions 1 \& 2 were selected following a quantitative analysis of
directions in the $(l^{II} = 0\arcdeg \rightarrow 180\arcdeg)$ and
$(l^{II} = 90\arcdeg \rightarrow 270\arcdeg)$ planes, as the optimal
directions in which to distinguish between the effects of the
normalization and the axis ratio of the spheroid. Bahcall and Soneira
(1980) explain why these planes are especially important for determining
the characteristics of the spheroid. We ran the model at $5\arcdeg$
intervals in $b^{II}$ in each of these planes, and calculated the
variation with $b^{II}$ of the ratio of spheroid counts to disk counts,
and the ratio of spheroid counts for pairs of near-degenerate models: for
example, using the notation (disk : spheroid\ normalization\ ratio,
spheroid\ axis\ ratio), $(800:1, 1.0):(500:1, 0.8)$. We then looked for
directions with a high spheroid : disk ratio (in order to minimize the
Poisson error in total spheroid counts as well as the error due to
contamination of spheroid counts by disk stars) while at the same time,
maximizing the quantity $\mid (800:1, 1.0):(500:1, 0.8) - 1\mid$.
Directions 1 \& 2 optimized both these quantities.

Directions 3--10 were selected because fields in these directions are
being prepared by the Digital Palomar Observatory Sky Survey (DPOSSII; S.
G. Djorgovski \& S. C. Odewahn 1998, private communication; Djorgovski et
al. 1997, 1998) for star count studies. It would then be possible to see
how the techniques discussed here in a theoretical sense work in practice.


\section{$\chi^2$ Analysis} \label{chisq}

The $\chi^2$ statistic is
\begin{equation} \label{chieq}
\chi^2 = \sum_{dir=1}^{N} \frac{\left[A_{model}(m_1,m_2,l,b)d\Omega -
A_{observed}(m_1,m_2,l,b)d\Omega \right]^2}{\sigma^2},
\end{equation}

where $dir=(l,b)$ is a given direction, $A(m_1,m_2,l,b)d\Omega$ is the
total number of spheroid stars with apparent magnitudes in the range $m_1
\leq m \leq m_2$ in the direction $(l,b)$ per projected sky area
$d\Omega$, and $\sigma$ is the error, which can have several components as
described below. Reduced $\chi^2$ is given by $\chi^2/\nu$ where $\nu$ is
the number of degrees of freedom, in this case equal to the number of
directions $N$ (i.e. the total predicted counts are {\it not} renormalized
to the same total counts as the data). A large value of reduced $\chi^2$
indicates that the null hypothesis that the $A_{observed}$'s are drawn
from the same population as the $A_{model}$'s is rather unlikely.

Since we do not currently have observed number counts in all the
directions, we picked one model out of the set covering the investigated
parameter-space to be the ``observed" number counts and compared it
against each of the rest of the models in the set (which represented the
``theory"). The process was then repeated till each of the models had been
picked as the one representing actual data.

We define the criterion for lifting the degeneracy of spheroid parameters
as obtaining the probability of finding a value of reduced $\chi^2$
greater than or equal to the ``observed" value $<$ 10$^{-5}$. That is, we
aim to reject the null hypothesis (that the ``observed" and ``theoretical"  
models are the same) at a confidence level of 0.001\%. The range of models
we used spanned the parameter space \((disk : spheroid\ normalization\
ratio) \times (spheroid\ axis\ ratio) = (500:1, 800:1)  \times (1.0, 0.8,
0.6, 0.4)\). Thus there are 56 possible pairings of models picked as
"observed" and "theoretical".

$\chi^2$ tests were carried out using the set of directions detailed in
\S~\ref{dirn} as ``bins." The directions yielding the highest contribution
to the $\chi^2$ value were noted as the optimal directions to use for
breaking the degeneracy. It was then checked whether $\chi^2$ tests using
just those optimal directions as bins would cause the degeneracy to be
lifted.

The error term $\sigma$ can have contributions from Poisson error,
magnitude error, and the systematic ``separation error'' which arises
because spheroid stars cannot be separated cleanly from old/thick disk
stars in observations, thereby introducing an error into the total
spheroid counts. In general, a magnitude error of order $\sim 0.1^{mag}$
contributes less to the error budget than the Poisson error. In a ``good''
direction, the separation error should contribute the least to the error
budget and the number counts are Poisson-noise-limited. In a ``bad''
direction, the separation error can be by far the greatest source of
uncertainty. In \S~\ref{nosep}, we first consider the case where the
spheroid counts {\it{can}} be cleanly separated, and then in
\S~\ref{sepanal} we consider the effect of adding the separation error.

\subsection{Analysis excluding separation error} \label{nosep}

$\chi^2$ tests were first carried out using the two-component (spheroid +
old disk) model, taking into account the Poisson error of the counts and a
magnitude error of 0.1$^{mag}$, but assuming that spheroid stars can be
separated cleanly from old/thick disk stars in observations.

The magnitude error was calculated as follows. Let $N(m_1 \leq m \leq
m_2)$ represent the total spheroidal number counts for apparent magnitude
range $m_1 \leq m \leq m_2$, and let the error in the zero-point be
$\delta m$. Then we calculate $N_+(m_1+\delta m \leq m \leq m_2+\delta m)$
and $N_-(m_1-\delta m \leq m \leq m_2-\delta m)$. The error in the total
spheroidal number counts due to the zero-point error in magnitude is then
given by $Max(\mid N-N_+ \mid, \mid N-N_- \mid)$.

\subsubsection{Directions 1 \& 2}

Directions 1 and 2 in Table~\ref{tbl-dir} were found to be the best
overall for discriminating between ordinarily degenerate models using
$\chi^2$ tests.

Using the notation $[(model\ taken\ as\ observation),\ (model\ taken\ as\
prediction)]$, almost all model pairings were distinguishable from each
other in the apparent magnitude range ${19 < V < 20}$. The exceptions were
$[(500:1,0.8),\ (800:1,1.0)]$ and $[(800:1,1.0),\ (500:1,0.8)]$. Upon
examining the range ${20 < V < 21}$ the degeneracy of these model pairings
were also lifted, while the other model pairings also remained
non-degenerate.

\subsubsection{Directions 3--10 $(19 < V < 20)$}

Considering the directions {3--10} which are being prepared by the DPOSSII
survey, $\chi^2$ tests in directions 3 and 4 in apparent magnitude range
${19 < V < 20}$ can distinguish between all model pairings with the
exceptions shown in Table~\ref{tbl-chi1}.

\placetable{tbl-chi1}

\subsubsection{Directions 3--10 $(20 < V < 21)$}

Considering the model pairings in the apparent magnitude range ${20 < V <
21}$ in the DPOSSII directions, $\chi^2$ tests can distinguish between all
model pairings in directions 3 and 4 except for those shown in
Table~\ref{tbl-chi2}.

\placetable{tbl-chi2}

\subsubsection{Discussion} \label{seperr}

Observationally, spheroid stars can be separated from old/thick disk stars
using the bi-modal distribution that appears in star color-frequency
profiles in certain directions at faint magnitude ranges. The blue and red
peaks consist of spheroid and disk stars, respectively. This distribution
has been interpreted as arising because the disk and spheroid components
have different density gradients in a magnitude-limited survey
(\cite{bs80}). The sharp density gradient in the disk for directions far
from the Galactic plane favors relatively nearby and intrinsically faint
(red) stars at faint apparent magnitudes. Conversely, the shallow density
gradient in the spheroid favors relatively distant, intrinsically bright
(blue) stars at the same faint apparent magnitudes, because the effective
volume increases at large distances.

$18 \leq V \leq 22$ is the apparent magnitude range where this
double-peaked distribution is most pronounced, enabling the cleanest
separation of spheroidal stars in observations (\cite{bs80},
\cite{rei93}). In this interval, the spheroid counts peak at $B-V \sim
0.5$, with a narrow range in color. This corresponds roughly to the
main-sequence turn-off, at $M_V \sim 4.5$. For a given apparent magnitude
interval $m_1 \leq m \leq m_2$, only stars within a range of absolute
magnitudes $M_1 \leq M \leq M_2$ contribute to the observed number counts.
The majority contribution to the blue spheroidal peak comes from stars
near the main sequence turn-off, with $+4 \leq M_V +6$ (corresponding to
$\sim 0.5 \rightarrow 1.0$ M$_\odot$). Evolved stars with $M_V \leq 3.5$
make negligible contribution to deep star counts at this magnitude range.

The shallow density gradient in the spheroid leads to a broad distribution
of number counts with distance in a given direction, peaking at $\sim 7$
kpc in the direction of the North Galactic Pole (\cite{bs80}). Therefore,
the fainter the magnitude range under consideration, the better the
performance of the $\chi^2$ statistic, because spheroid number counts are
still rising at $V \sim 22$. Thus the Poisson errors and magnitude errors
at fainter magnitudes are smaller. This is the reason that the magnitude
range $20 \leq V \leq 21$ is more useful for lifting degeneracies using
the $\chi^2$ tests than the brighter range $19 \leq V \leq 20$. Once we go
even fainter to the range $21 \leq V \leq 22$, photometric uncertainties
and observational errors, such as star-galaxy separation and contamination
of the blue spheroidal population by quasars and compact emission line
galaxies (CELGs) increasingly start to affect counts. Thus the
intermediate apparent magnitude range $20 \leq V \leq 21$ is the optimal
one to use for the purpose of lifting degeneracies by this method.

\subsection{Analysis including separation error} \label{sepanal}

\begin{figure} 
\plotone{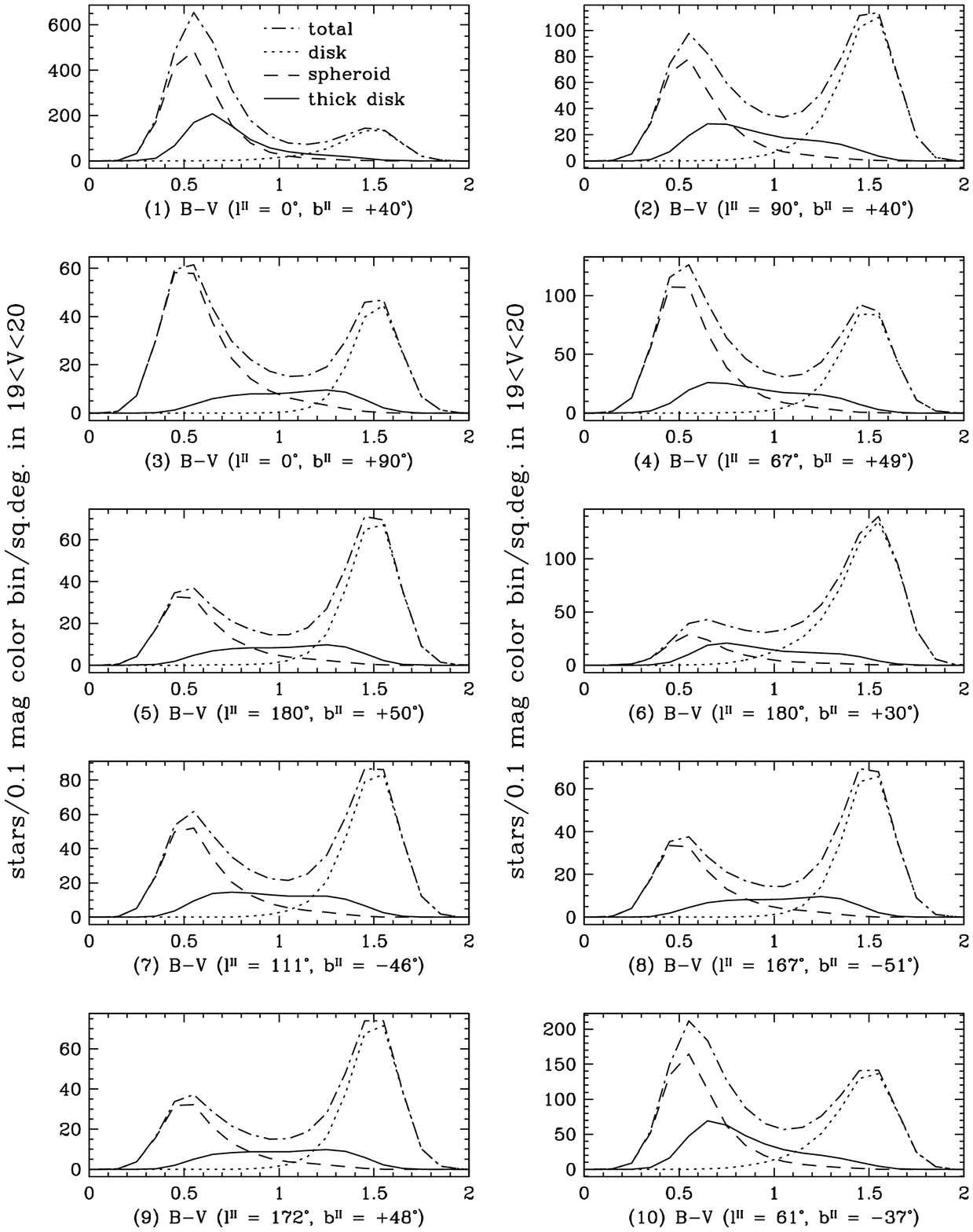} 
\caption{Frequency distribution of star colors for a ``standard" model
$(500:1,0.8)$ (\cite{bah86}) plus a thick-disk as described in
\S~\ref{model}, in the apparent magnitude range $19 < V < 20$.
\label{fig-19}}
\end{figure}

\begin{figure}
\plotone{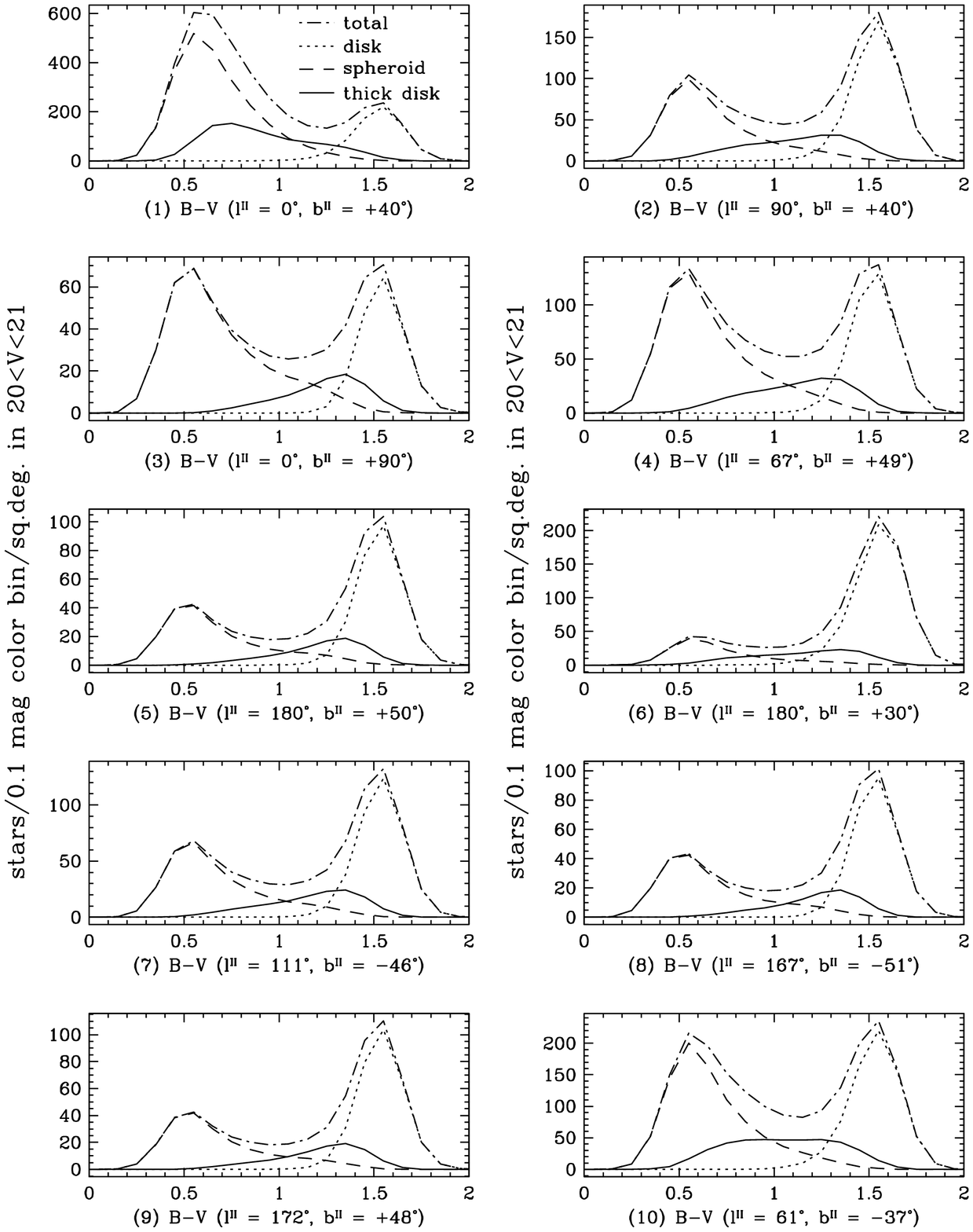}
\caption{Frequency distribution of star colors for a ``standard" model
$(500:1,0.8)$ (\cite{bah86}) plus a thick-disk as described in
\S~\ref{model}, in the apparent magnitude range $20 < V < 21$.
\label{fig-20}}
\end{figure}

Figures~\ref{fig-19} and~\ref{fig-20} show frequency distributions of star
colors for a particular ``standard" model, in the ten directions given in
Table~\ref{tbl-dir}, for magnitude ranges ${19 < V < 20}$ and ${20 < V <
21}$. The maximum error in separating out the spheroid stars arises if
there is a thick disk component which tends to fill up the valley between
the double-peaks due to spheroid and old disk stars, or in the worst
cases, peaks within the blue spheroid peak. The separation error is
calculated as the area under the blue peak in total (spheroid + old disk +
thick disk) counts, minus the calculated (i.e. model) spheroid counts. The
``blue peak'' is taken to be bounded on the red side by the color
corresponding to the minimum point of the total counts.

\placefigure{fig-19}
\placefigure{fig-20}

Now, we consider the effect of including the systematic error due to
separating spheroid stars from old/thick disk stars in observations, using
the three-component (spheroid + old disk + thick disk) model.  The
separation error is an overestimate since this contamination in
observations can be corrected for to some extent by subtracting model disk
counts from the total counts.

The contamination due to the thick disk was found to dominate that due to
old disk stars in all cases. This error was found to be smaller in the
apparent magnitude range ${20 < V < 21}$ than in the range ${19 < V <
20}$. Hence the following analysis was only carried out for the fainter
range. Further, since the ratio of separation error to total spheroid
counts increases significantly for models with smaller axis ratios, the
models with axis ratio 0.4 were excluded from the analysis. Thus, model
parameter space was restricted to \((500:1, 800:1) \times (1.0, 0.8,
0.6)\).

Upon including the separation error, directions 1 and 2, found to be the
best overall in discriminating between model parameters, were found to be
virtually useless, since the thick disk counts peaked within the spheroid
peak for these directions.

In the case of the DPOSSII directions (3--10), all the model pairings
which could be distinguished in this apparent magnitude range using
$\chi^2$ tests in directions 3 and 4 remained non-degenerate, except
$[(800:1,0.6),\ (500:1,0.6)]$, which became degenerate. The rest of the
model pairings which had been non-degenerate in directions given in
Table~\ref{tbl-chi2} also became degenerate.

Table~\ref{tbl-chi3} shows whether these degenerate models can be
discriminated using other combinations of directions.

\placetable{tbl-chi3}

During this analysis, it was found directions 7 and 9 are generally good
for minimizing the effects due to the thick disk. Conversely, directions 1
and 10 are particularly good for looking for the presence of the thick
disk.

\section{Conclusions} \label{finish}

We found that directions 1 ($l^{II} = 0\arcdeg, b^{II} = 40\arcdeg$) and 2
($l^{II} = 90\arcdeg, b^{II} = 40\arcdeg$) in apparent magnitude range $20
< V < 21$ were the most effective in distinguishing between effects due to
degenerate parameters, assuming that spheroid stars could be separated
cleanly from disk and thick-disk stars.

In the case that the maximum possible separation error (i.e. uncorrected
for disk and thick disk counts) was included in the analysis, directions 3
(the North Galactic Pole) and 4 ($l^{II} = 67\arcdeg, b^{II} = 49\arcdeg$)  
in apparent magnitude range $20 < V < 21$ were the most useful. In those
models where these directions failed to lift the degeneracy, directions 7
($l^{II} = 111\arcdeg, b^{II} = -46\arcdeg$) and 9 ($l^{II} = 172\arcdeg,
b^{II} = 48\arcdeg$) were the best for minimizing contamination from the
thick disk, which dominates the separation error.

Directions 1 ($l^{II} = 0\arcdeg, b^{II} = 40\arcdeg$) and 10 ($l^{II} =
61\arcdeg, b^{II} = -37\arcdeg$) were found to be the most effective in
detecting the presence of a thick disk.

This is an exciting time in which deep, high-quality data-sets covering
large areas of the sky are about to become available from much-anticipated
surveys such as the Sloan Digital Sky Survey (SDSS, see e.g. \cite{gun93},
\cite{gun95}) and the DPOSSII (see \S~\ref{dirn} for references). The use
of survey data for star count studies is complicated by the fact that many
existing ``standard" Galactic structure models are based on
color-magnitude diagrams (CMDs) and luminosity functions (LFs) determined
in the standard Johnson-Morgan-Cousins $UBV$ photometric system, and are
only useful for comparison with data in the visual band or in photometric
systems to which accurate transformations exist.  However, in cases such
as that of the SDSS, which uses the non-standard $u'g'r'i'z'$ photometric
system (\cite{fuk96}), such models will soon be re-created using LFs and
CMDs determined in the SDSS photometric bands. The DPOSSII uses three
photographic $JFN$ bands calibrated to the Gunn $gri$ bands. Though these
are different from the photoelectric/CCD $gri$ bands (Palomar, 4-Shooter,
SDSS) they are well-defined. Once the work of creating a model based on
the SDSS bands is completed, it should be fairly straightforward to
translate the stellar sequences and LFs to the DPOSSII $gri$ bands. Until
such models are available, a different approach to star count models, such
as the evolutionary stellar population synthesis technique (\cite{fan99})
can be used.

The possibility of fitting models simultaneously in a multitude of
directions with large samples containing number, magnitude and color
information should revolutionize studies of Galactic structure using star
counts.


\acknowledgements

The author is very grateful to John Bahcall for his helpful advice. She wishes
to thank Andy Gould for providing machine-readable versions of Figure 2 of 
Gould et al.\ (1997) and Figure 3 of Gould et al.\ (1998); David Spergel and 
Xiaohui Fan for instructive discussions on Galactic Structure models; and Mark Jackson for useful comments on the manuscript. 


\clearpage
\begin{deluxetable}{ll}
\footnotesize
\tablecaption{The Assumed Stellar Distributions. \label{tbl-dist}}
\tablewidth{0pt}
\tablehead{
\colhead{Component} & \colhead{Distribution}
}
\startdata
Disk & $n_D = n_D (R_0) \exp [-z/H(M_V)] \exp [-(x-R_0)/h]$ \nl \nl
Spheroid & $n_{sph} = 
n_{sph}(R_0)(R/R_0)^{-7/8}{ \exp [-10.093(R/R_0)^{1/4} + 10.093]}$ \nl
& $\times 125(R/R_0)^{-6/8} { \exp [-10.093(R/R_0)^{-1/4} + 10.093]}, 
\ R < 0.03 R_0$ \nl
& $\times [1-0.08669/(R/R_0)^{1/4}],\ R \geq 0.03 R_0$ \nl
\nl
Thick Disk & $n_{TD} = n_{TD} (R_0) \exp [-z/H_{TD}] \exp [-(x-R_0)/h]$ \nl
\nl
Normalization & $n_D(R_0) = 0.13 pc^{-3}, \ n_{TD}(R_0) = 0.0026 pc^{-3},$ \nl
& $n_{sph}(R_0) = 0.00026 pc^{-3}$ or $0.0001625 pc^{-3}$ depending on model 
\nl
\enddata
\tablecomments{ Here $z$ is the distance perpendicular to the plane, $x$
is the galactocentric distance in the plane, and $h$ is the old disk
scale length. Galactocentric distance $R = (x^2 + z^2/ \kappa^2)^{1/2}$,
where $\kappa$ is the axis ratio and $1-\kappa$ is the ellipticity. We
adopt $R_0 = 8$ kpc and $h = 3.5$ kpc. The old disk scale height $H(M_V)$
is given in Bahcall and Soneira (1980). The thick disk is taken to have a
scale height $H_{TD} = 1.2$ kpc and a 47-Tuc-like CMD.}
\end{deluxetable}


\clearpage
\begin{deluxetable}{cr@{}lr@{}ll }
\footnotesize
\tablecaption{A Representative Set of Directions. \label{tbl-dir}}
\tablewidth{0pt}
\tablehead{
\colhead{ Direction } & \colhead{$l^{II}$(1950)} & \colhead{} & 
\colhead{$b^{II}$(1950)} & \colhead{} &
\multicolumn{1}{c}{Comments\tablenotemark{a}}
}
\startdata
 1..... &  0	   & \arcdeg & +40 & \arcdeg & BS9\tablenotemark{b} \nl
 2..... &  90      && +40 && Between BS4 and BS7\tablenotemark{b}\nl
 3..... &  \nodata && +90 && Galactic pole (SA57, BS1)\tablenotemark{c,d} \nl
 4..... &  67 	   && +49 && BS16\tablenotemark{c,d} \nl
 5..... & 180      && +50 && BS6\tablenotemark{c} \nl
 6..... & 180      && +30 && BS8\tablenotemark{c} \nl
 7..... & 111      && --$\,$46 && BS11, SA68\tablenotemark{c} \nl
 8..... & 167      && --$\,$51 && BS13\tablenotemark{c} \nl
 9..... & 172      && +48 && BS14\tablenotemark{c} \nl
10..... &  61      && --$\,$37 && BS17\tablenotemark{c} \nl
\enddata
\tablenotetext{a}{The field identification BS{\it n} denotes the
designation assigned to a given direction in Table 1 of Bahcall and
Soneira (1981), where quantities that can be efficiently studied in that
direction are listed. The Selected Area number is also given for fields
which have been studied.}  
\tablenotetext{b}{These directions were determined to be the best for 
distinguishing between effects due to spheroid flattening and
normalization using the two-component (old disk + spheroid) standard
model.}
\tablenotetext{c}{These fields are being made available by the DPOSSII
survey for star count studies.}
\tablenotetext{d}{These directions were determined to be the best for 
determining the spheroid parameters out of the DPOSSII directions.}
\end{deluxetable}


\clearpage

\begin{deluxetable}{cc}
\footnotesize
\tablecaption{Model pairings degenerate in directions 3 and 4 for ${19 < V < 
20}$ excluding separation error. \label{tbl-chi1}}
\tablewidth{0pt}
\tablehead{
\colhead{Model pairing\tablenotemark{a}} & \colhead{Min. directions required to 
lift degeneracy}
}
\startdata
$[(500:1,0.8),\ (800:1,1.0)]$ 	&	6, 7		\nl
$[(800:1,1.0),\ (500:1,0.8)]$	&	6, 7, 10	\nl
$[(500:1,0.6),\ (800:1,0.8)]$   &         \nodata\tablenotemark{b}	
	\nl
$[(800:1,0.8),\ (500:1,0.6)]$ 	&	\nodata\tablenotemark{b}	
	\nl
$[(500:1,0.4),\ (800:1,0.6)]$	&	\nodata\tablenotemark{b}		\nl
$[(800:1,0.4),\ (500:1,0.4)]$	&	4, 10		\nl
\enddata
\tablenotetext{a}{Uses notation {\it [(model\ taken\ as\ observation),\ (model\ 
taken\ as\
prediction)]}.}
\tablenotetext{b}{This model pairing cannot be distinguished using the DPOSSII
directions 3--10 in the apparent magnitude range ${19 < V < 20}$.} 
\end{deluxetable}


\clearpage

\begin{deluxetable}{cc}
\footnotesize
\tablecaption{Model pairings degenerate in directions 3 and 4 for ${20 < V < 
21}$ excluding separation error. \label{tbl-chi2}}
\tablewidth{0pt}
\tablehead{
\colhead{Model pairing\tablenotemark{a}} & \colhead{Min. directions required to 
lift degeneracy}
}
\startdata
$[(500:1,0.8),\ (800:1,1.0)]$       &        6, 10 \nl
$[(800:1,1.0),\ (500:1,0.8)]$       &        6, 10 \nl
$[(500:1,0.6),\ (800:1,0.8)]$       &        3, 6  \nl
$[(800:1,0.8),\ (500:1,0.6)]$       &        \nodata\tablenotemark{b}   \nl
\enddata
\tablenotetext{a}{Uses notation {\it [(model\ taken\ as\ observation),\ (model\ 
taken\ as\ prediction)]}.}
\tablenotetext{b}{This model pairing cannot be distinguished using the DPOSSII
directions 3--10 in the apparent magnitude range ${20 < V < 21}$.} 
\end{deluxetable}


\clearpage

\begin{deluxetable}{cc}
\footnotesize
\tablecaption{Model pairings degenerate in directions 3 and 4 for ${20 < V < 
21}$ 
including separation error. \label{tbl-chi3}}
\tablewidth{0pt}
\tablehead{
\colhead{Model pairing\tablenotemark{a}} & \colhead{Min. directions required to 
lift degeneracy}
}
\startdata
$[(500:1,0.8),\ (800:1,1.0)]$     &          7, 9	\nl
$[(800:1,1.0),\ (500:1,0.8)]$     &          \nodata\tablenotemark{b}	\nl
$[(500:1,0.6),\ (800:1,0.8)]$     &          \nodata\tablenotemark{b} 	\nl
$[(800:1,0.8),\ (500:1,0.6)]$     &          \nodata\tablenotemark{b}	\nl
$[(800:1,0.6),\ (500:1,0.6)]$     &          7, 9	\nl
\enddata
\tablenotetext{a}{Uses notation {\it [(model\ taken\ as\ observation),\ (model\ 
taken\ as\ prediction)]}.}
\tablenotetext{b}{This model pairing cannot be distinguished using the DPOSSII
directions 3--10 in the apparent magnitude range ${20 < V < 21}$ using
$\chi^2$ tests including the separation error.}
\end{deluxetable}


\clearpage

\clearpage

\clearpage

\clearpage
\begin{thebibliography}{}
\bibitem[Bahcall \& Soneira\ 1980]{bs80} Bahcall, J. N., \& Soneira, R. M.
1980, \apjs, 44, 73
\bibitem[Bahcall \& Soneira\ 1981]{bs81} Bahcall, J. N., \& Soneira, R.
M. 1981, \apjs, 47, 357
\bibitem[Bahcall \& Soneira\ 1984]{bs84} Bahcall, J. N., \& Soneira, R.M.
1984, \apjs, 55, 67
\bibitem[Bahcall\ 1986]{bah86} Bahcall, J. N. 1986, \araa, 24, 577
\bibitem[Dahn et al.\ 1995]{dah95} Dahn, C. C., Liebert, J. W., Harris, H., \& Guetter, H. C. 1995, ESO Workshop on the Bottom of the Main Sequence and Beyond, ed. C. G. Tinney (Heidelberg: Springer), 239
\bibitem[Djorgovski et al.\ 1997]{djo97} Djorgovski, S. G., de Carvalho, R.
R., Odewahn, S. C., Gal, R. R., Roden, J., Storolz, P., \& Gray, A. 1997,
Applications of Digital Image Processing XX, ed. Tescher, proc.
S.P.I.E. 3164
\bibitem[Djorgovski et al.\ 1998]{djo98} Djorgovski, S. G., Gal, R. R.,
Odewahn, S. C., de Carvalho, R. R., Brunner, R., Longo, G., \&
Scaramella, R. 1998, to appear in Wide Field Surveys in Cosmology,
ed. Colombi \& Mellier, proc. XIV IAP Colloq., in press.
\bibitem[Fan 1999]{fan99} Fan, X. 1999, \aj, 117, 2528
\bibitem[Fukugita et al. 1996]{fuk96} Fukugita, M., Ichikawa, T., Gunn, J.
E., Doi, M., Shimasaku, K., \& Schneider, D. P. 1996, \aj, 111(4), 1748
\bibitem[Gilmore 1984]{gil84} Gilmore, G.F. 1984, \mnras, 207, 223
\bibitem[Gould et al.\ 1997]{gou97} Gould, A., Bahcall, J. N., \& Flynn, C.  
1997, \apj, 482, 913
\bibitem[Gould et al.\ 1998]{gou98} Gould, A., Flynn, C., \& Bahcall, J. N.  
1998, \apj, 503, 798
\bibitem[Gunn \& Knapp 1993]{gun93} Gunn, J. E., \& Knapp, G. R. 1993, in
ASP Conf. Ser. 43, Sky Surveys: Protostars to Protogalaxies, ed. Soifer, 267
\bibitem[Gunn \& Weinberg 1995]{gun95} Gunn, J. E., \& Weinberg, D. H.
1995, in Wide Field Spectroscopy and the Distant Universe, ed.
Maddox \& Ara\`g on-Salamanca, (World Scientific, Singapore)
\bibitem[Press et al. 1992]{pre92} Press, W. H., Teukolsky, S. A.,
Vetterling, W. T., \& Flannery, B. P. 1992, Numerical Recipes in C
(2$^{nd}$ Edition; C.U.P.)
\bibitem[Reid \& Majewski\ 1993]{rei93} Reid, I. N., \& Majewski, S.R.
1993, \apj, 409, 635
\bibitem[Robin \& Cr\'ez\'e 1986a]{rob86a} Robin, A., \& Cr\'ez\'e, M.
1986a, A\&A, 157, 1
\bibitem[Robin \& Cr\'ez\'e 1986b]{rob86b} Robin, A., \& Cr\'ez\'e, M.    
1986b, A\&AS, 64, 53
\end{thebibliography}
\end{document}